\documentclass[a4paper]{article}
\usepackage{footmisc}
\pdfoutput=1
\usepackage[latin9]{inputenc}
\usepackage{array}
\usepackage{float}
\usepackage{multirow}
\usepackage{amsmath}
\usepackage{amssymb}
\usepackage{graphicx}
\usepackage{subfig}
\usepackage{cancel}
\usepackage{xcolor}
\usepackage[colorlinks=true,linkcolor=red]{hyperref} 
\usepackage{eso-pic} 

\makeatletter

\pdfpageheight\paperheight
\pdfpagewidth\paperwidth

\providecommand{\tabularnewline}{\\}

\usepackage{spconf}

\usepackage{bm}

\ninept

\title{Language Model Integration based on Memory Control for \\Sequence to sequence Speech Recognition}
\name{\parbox{1.0\linewidth}{\center
Jaejin Cho$^{1*}$, Shinji Watanabe$^{1*}$, Takaaki Hori$^{2*}$, Murali Karthick Baskar$^{3*}$,
\\Hirofumi Inaguma$^{4*}$, Jesus Villalba$^{1*}$, Najim Dehak$^{1*}$\thanks{* The affiliations of the authors may differ from their current affiliations.}}}
 
\address{
$^1$Johns Hopkins University, $^2$Mitsubishi Electric Research Laboratories (MERL),
\\$^3$Brno University of Technology, $^4$Kyoto University\\
{\small\tt youjojo8478@gmail.com}
}

\makeatother

\begin{document}
\ninept
\AddToShipoutPictureBG*{%
  \AtPageUpperLeft{%
    \hspace{1cm}
    \raisebox{-1cm}{
      \fbox{\hyperref[sec:notice]{\Large \textbf{Important Notice: Please read this note before reading/citing this paper.}}}%
    }%
  }%
}

\maketitle

\begin{abstract}

In this paper, we explore several new schemes to train a seq2seq model to 
integrate a pre-trained LM. 
Our proposed fusion methods focus on the memory cell state and the hidden state in the seq2seq decoder long short-term memory (LSTM), and the memory cell state is updated by the LM unlike the prior studies.
This means the memory retained by the main seq2seq would be adjusted by the external LM. 
These fusion methods have several variants depending on the architecture of this memory cell update and the use of memory cell and hidden states which directly affects the final label inference.
We performed the experiments to show the effectiveness of the proposed methods in a mono-lingual ASR setup on the Librispeech corpus and in a transfer learning setup from a multilingual ASR (MLASR) base model to a low-resourced language. In Librispeech, our best model improved WER by 3.7\%, 2.4\% for test clean, test other relatively to the shallow fusion baseline, with multi-level decoding. In transfer learning from an MLASR base model to the IARPA Babel Swahili model, the best scheme improved the transferred model on eval set by 9.9\%, 9.8\% in CER, WER relatively to the 2-stage transfer baseline.
\end{abstract}


\section{introduction}

As deep learning prospers in most research fields, systems based on it keep improving, and become the state-of-the-art in most of the scenarios. The sequence to sequence (seq2seq) model is one of the kind that heavily depends on deep learning techniques, and it is used in many sequence mapping problems such as automatic speech recognition (ASR)~\cite{graves2014towards,chorowski2014end,bahdanau2016end,chan2016LAS} and machine translation~\cite{gulcehre2015firstSFnDF,luong2015effective,bahdanau2014neural}. In~\cite{chan2016LAS}, a seq2seq model with attention mechanism is introduced in ASR. Though the performance lagged behind highly-optimized conventional systems, e.g. CLDNN HMM system~\cite{sainath2015convolutional}, it enabled to map a sequence of feature vectors to a sequence of characters, 
with a single neural network, in an end-to-end manner. In~\cite{watanabe2017attctc}, the authors apply a multi-task learning scheme to train an attentional seq2seq model with connectionist temporal classification (CTC) objective function~\cite{graves2014towards,graves2006connectionist} as auxiliary loss. Adding the CTC loss to train the model reduces the burden of the attention model to learn monotonic attention.

 In the seq2seq ASR setup, the language model (LM) takes an important role as it is already shown in hybrid ASR systems~\cite{dahl2012context,bahl1989tree}. 
 However, compared to the conventional ASR, there have been only a few studies on ways to integrate an LM into seq2seq ASR~\cite{hori-interspeech-2017,graves2012sequence,miao2015eesen}. In this direction, the authors in~\cite{gulcehre2015firstSFnDF} introduce two methods integrating an LM into a decoder of the end-to-end neural machine translation (NMT) system. First method was \textit{shallow fusion} where the model decodes based on a simple weighted sum of NMT model and recurrent neural network LM~\cite{mikolov2010recurrent} (RNNLM) scores. The next one was called \textit{deep fusion} where they combine a mono-lingual RNNLM with a NMT model by learning parameters that connect hidden states of a separately trained NMT model and RNNLM. While the parameters connecting the hidden states are trained, parameters in NMT and RNNLM are frozen. Recently in ASR research, a scheme called \textit{cold fusion} was introduced, which trains a seq2seq model from scratch in assistance with a pre-trained RNNLM~\cite{sriram2017coldfusion}. In contrast to the previous methods, the parameters of the seq2seq model are not frozen during training although pre-trained RNNLM parameters are still kept frozen. The results showed the model trained this way outperforms \textit{deep fusion} in decoding as well as reducing the amount of data in domain adaptation. Later, more experiments were done comparing all three methods~\cite{toshniwal2018comparisonSFnDFnCF}. In the paper, they observe that \textit{cold fusion} works best among all three methods in the second pass re-scoring with a large and production-scale LM. The previous research has shown the potential of training a seq2seq model utilizing a pre-trained LM. However, it seems only effective to limited scenarios such as domain adaptation and second-pass re-scoring. Thus, studies on better ways of integrating both models need to be explored.

 In this paper, we explored several new fusion schemes to train a seq2seq model jointly with a pre-trained LM. Among them, we found one method that works consistently better than other fusion methods over more general scenarios. The proposed methods focus on updating the memory cell state as well as the hidden state of the seq2seq decoder long short-term memory (LSTM)~\cite{hochreiter1997long}, given the LM logit or hidden state. This means that the memory retained by the main seq2seq model will be adjusted by the external LM for better prediction. The fusion methods have several variants according to the architecture of this memory cell update and the use of memory cell and hidden states, which directly affects the final label inference.
 Note that we used LSTM networks for RNNs throughout whole explanations and experiments. The proposed methods, however, can be  applied to different variant RNNs such as gated recurrent unit (GRU)~\cite{chung2014empirical} only with minimal modification.

 The organization of this paper is as follows. First, we describe previous fusion methods as a background in Section~\ref{ses:background}. Then, in Section~\ref{ses:proposed}, we explain our proposed methods in detail. Experiments with previous and proposed methods are presented in Section~\ref{ses:exp}. Lastly, we conclude the paper in Section~\ref{ses:conclusion}.
  
\section{Background: Shallow fusion, Deep fusion, and Cold fusion in ASR} \label{ses:background}

\subsection{Shallow fusion}
 In this paper, we denotes as {\it shallow fusion} a decoding method based on the following convex score combination of a seq2seq model and LM during beam search,
 \begin{equation}
 \label{eq:shallow_comb}
 \hat{y} = \underset{y}{\operatorname{argmax}}\,(\log p(y|x) + \gamma \log p(y))
 \end{equation}
 where $x$ is an input acoustic frame sequence while $\hat{y}$ is the predicted label sequence selected among all possible $y$.
 The predicted label sequence can be a sequence of characters, sub-words, or words, and this paper deals with character-level sequences.
 $\log p(y|x)$ is calculated from the seq2seq model and $\log p(y)$ is calculated from the RNNLM. 
 Both models are separately trained but their scores are combined in a decoding phase. 
 $\gamma$ is a scaling factor between 0 and 1 that needs to be tuned manually.
 
\subsection{Deep fusion}
In \textit{deep fusion}, the seq2seq model and RNNLM are combined with learnable parameters. 
Two models are first trained separately as in \textit{shallow fusion}, and then both are frozen while the connecting linear transformation parameters, i.e. $\bm{v}$, $b$, $\bm{W}$, and $\bm{b}$ below, are trained.
\begin{subequations}
\begin{equation}
\label{eq:df_1}
    g_t = \sigma(\bm{v} ^{\text{T}} \bm{s}_t^{\text{LM}} + b)
\end{equation}
\begin{equation}
\label{eq:df_2}
    \bm{s}_t^{\text{DF}} = [\bm{s}_t; g_t \bm{s}_t^{\text{LM}}]
\end{equation}
\begin{equation}
    \hat{p}(y_{t}|y_{<t},x) = \text{softmax}(\bm{W} \bm{s}_t^{\text{DF}}+\bm{b})
\end{equation}
\end{subequations}
$\bm{s}_t^{\text{LM}}$ and $\bm{s}_t$ are hidden states at time $t$ from an RNNLM and a decoder of the seq2seq model, respectively.
$\sigma(\cdot)$ in Eq.~\eqref{eq:df_1} is the sigmoid function to generate $g_t$, which acts as a scalar gating function in Eq.~\eqref{eq:df_2} to control the contribution of $\bm{s}_t^{\text{LM}}$ to the final inference.

\subsection{Cold fusion}
In contrast to the two previous methods, \textit{cold fusion} uses the RNNLM in a seq2seq model training phase.
The seq2seq model is trained from scratch with the pre-trained RNNLM model whose parameters are frozen and learnable parameters $\bm{W}_k$ and $\bm{b}_k$, where $k=1,2,3$. The equation follows
\begin{subequations} \label{eq:cf}
\begin{equation}
    \bm{h}_{t}^\text{LM} = \bm{W}_1 \bm{l}_{t}^\text{LM} + \bm{b}_1
\end{equation}
\begin{equation}
    \bm{g}_{t} = \sigma(\bm{W}_2[\bm{s}_{t}; \bm{h}_{t}^\text{LM}] + \bm{b}_2)
\end{equation}
\begin{equation}
\label{eq:cf_3}
    \bm{s}_{t}^\text{CF} = [\bm{s}_{t};\bm{g}_{t}\odot \bm{h}_{t}^\text{LM}]
\end{equation}
\begin{equation}
    \hat{p}(y_{t}|y_{<t}, x) = \text{softmax}(\text{ReLU}(\bm{W}_3 \bm{s}_{t}^\text{CF} + \bm{b}_3))
\end{equation}
\end{subequations}
where $\bm{l}_t^{\text{LM}}$ is the logit at time step $t$ from the RNNLM.
As opposed to a scalar $g_t$ in Eq.~\eqref{eq:df_2} of \textit{deep fusion}, it is now a vector $\bm{g}_t$ in \textit{cold fusion}, meaning the gating function is applied element-wise, where $\odot$ means element-wise multiplication.
$\text{ReLU}(\cdot)$ is a rectified linear function applied element-wise. Applying it before $\text{softmax}(\cdot)$ was shown to be helpful empirically in~\cite{sriram2017coldfusion}.

\begin{figure}[ht!]
\centering    
\subfloat[cold fusion]{\includegraphics[width=0.5\textwidth, keepaspectratio]{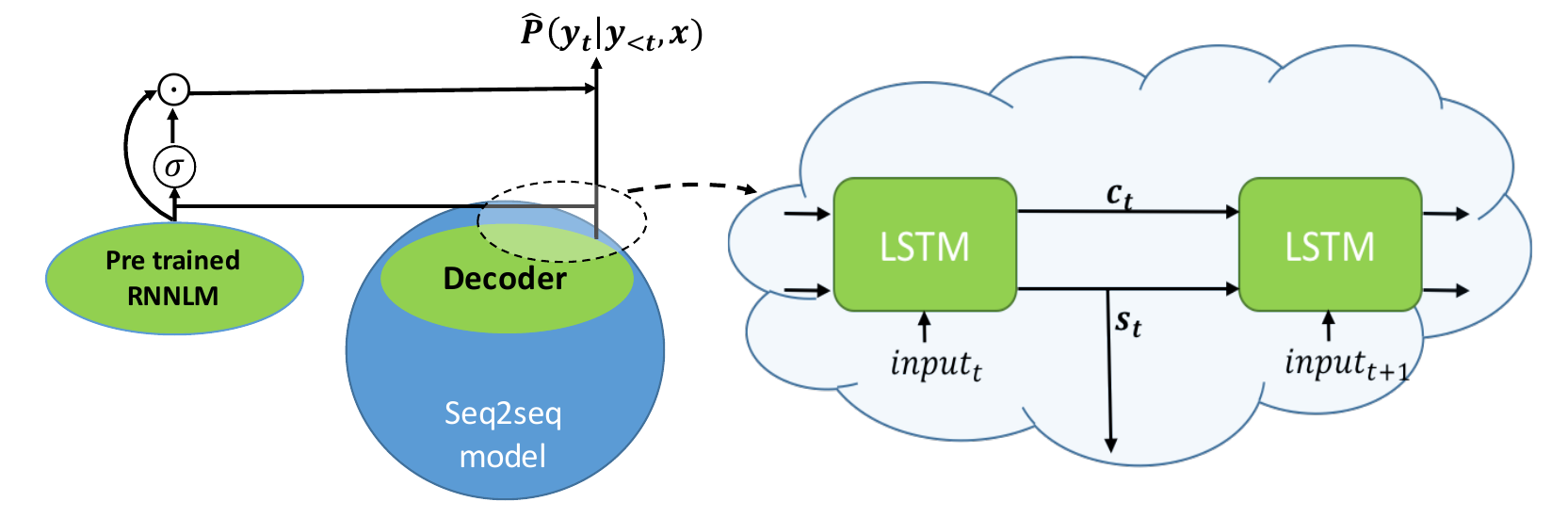}\label{fig:subfig5}}    

\subfloat[cell update]{\includegraphics[width=0.5\textwidth, keepaspectratio]{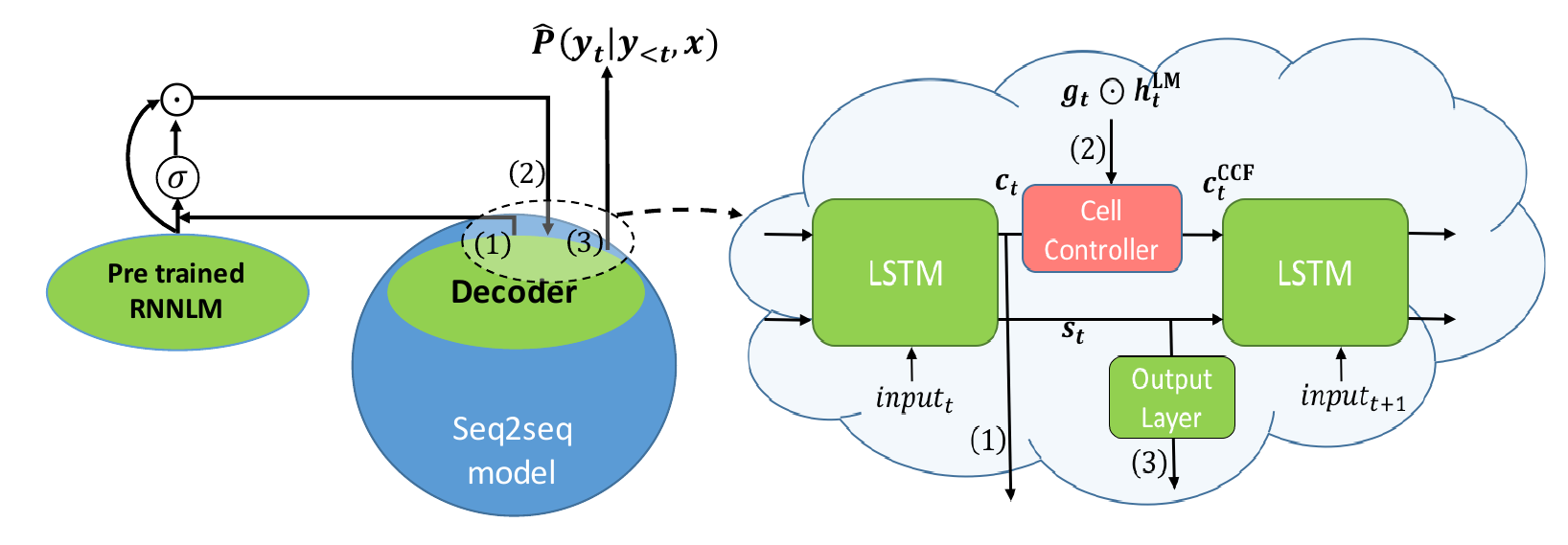}\label{fig:subfig6}}    

\subfloat[cell and state update + cold fusion]{\includegraphics[width=0.5\textwidth, keepaspectratio]{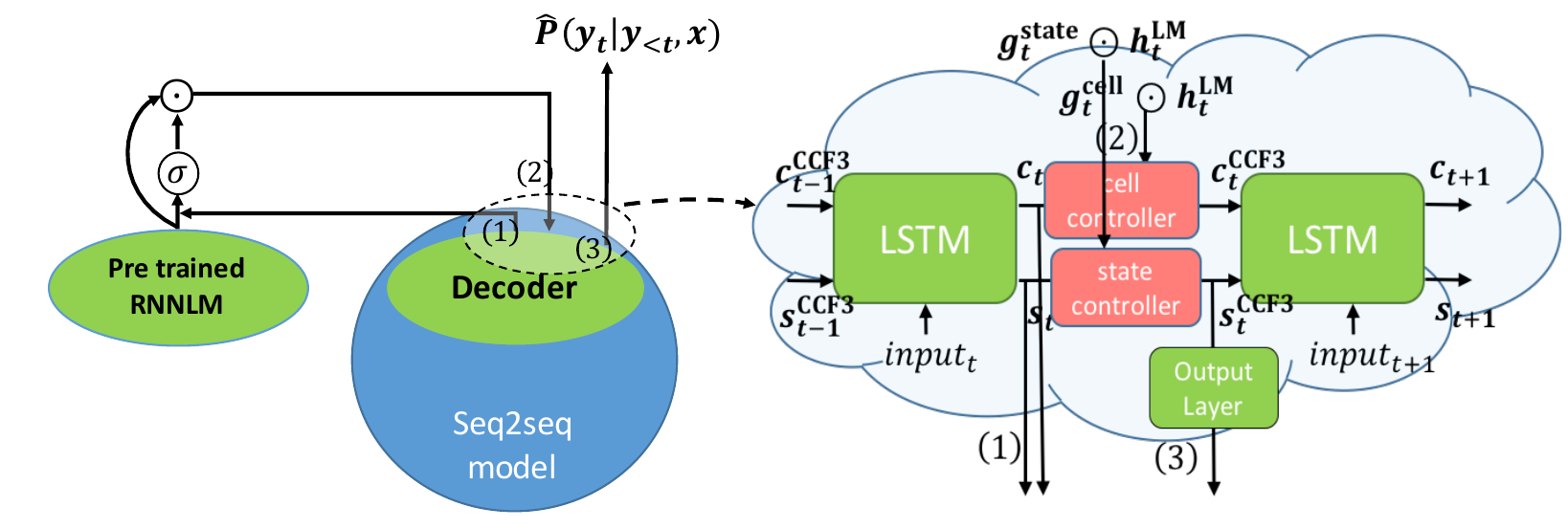}\label{fig:subfig7}}    
\caption[Optional caption for list of figures 5-8]{High-level illustration of fusion methods in training. (b) and (c) correspond to ~\textit{cell control fusion 1} and ~\textit{cell control fusion 3} respectively among our proposed methods. Output layer here is a linear transformation, with non-linearity only when having it benefits empirically}   
\label{fig:highlevpic}
\end{figure}

\section{Proposed methods} \label{ses:proposed}

In this paper, we propose several fusion schemes to train a seq2seq model well integrated with a pre-trained RNNLM.
We mainly focus on updating hidden/memory cell states in the seq2seq LSTM decoder given the LM logit/hidden state.
The first proposed method uses the LM information to adjust the memory cell state of the seq2seq decoder.
Then, the updated cell state replaces the original input cell state of the LSTM decoder to calculate states at the next time step.
This fusion scheme is inspired by \textit{cold fusion}, but they differ in that the new method directly affects the memory cell maintaining mechanism in the LSTM decoder to consider the linguistic context obtained from the LM.
Then, we further extend this idea with multiple schemes that use the LM information not only for updating the memory cell states but also hidden states in the seq2seq LSTM decoder, which further affect the final inference and attention calculation directly.
Figure~\ref{fig:highlevpic} visualizes the schemes in high-level to help understanding.

\subsection{LM fusion by controlling memory cell state} \label{point1}
In Eq. \eqref{eq:cf_3} at \textit{cold fusion}, the gated RNNLM information, $\bm{g}_{t}\odot \bm{h}_{t}^\text{LM}$, is concatenated with the decoder hidden state, $\bm{s}_t$. Then, the fused state, $\bm{s}_t^{\text{CF}}$ is used to predict the distribution of the next character.
However, the gated RNNLM information can be used in a different way to update the cell state in the seq2seq decoder, as in Eq. \eqref{eq:ccf_cell}.
\begin{subequations} \label{eq:ccf}
\begin{equation} \label{eq:ccf_logit_transf}
    \bm{h}_{t}^\text{LM} = \tanh(\bm{W}_1 \bm{l}_{t}^\text{LM} + \bm{b}_1)
\end{equation}
\begin{equation} \label{eq:ccf_gate}
    \bm{g}_{t} = \sigma(\bm{W}_2[\bm{c}_t; \bm{h}_{t}^\text{LM}] + \bm{b}_2)
\end{equation}
\begin{equation} \label{eq:ccf_cell}
    \bm{c}_{t}^\text{CCF} = \bm{c}_{t} + \bm{g}_{t}\odot \bm{h}_{t}^\text{LM}
\end{equation}
\begin{equation} \label{eq:ccf_dec}
    \bm{s}_{t+1}, \bm{c}_{t+1} = \text{LSTM}(\text{input}, \bm{s}_t, \bm{c}_t^\text{CCF})
\end{equation}
\begin{equation} \label{eq:ccf_inference}
     \hat{p}(y_{t}|y_{<t}, x) = \text{softmax}(\bm{W}_3 \bm{s}_t + \bm{b}_3)\;.
\end{equation}
\end{subequations}
In this method, we add the gated RNNLM information to the original cell state, $\bm{c}_t$. $\text{LSTM}(\cdot)$ in Eq.~\eqref{eq:ccf_dec} is the standard LSTM function, which takes previous cell and hidden states and an input came from an attention context vector, and updates the cell and hidden states for the next time step.
In our case, when the LSTM decoder updates its cell state, it uses $\bm{c}_{t}^\text{CCF}$ instead of $\bm{c}_{t}$, which contains additional linguistic context obtained from an external LM. We call this fusion as \textit{cell control fusion 1} throughout the paper. Here, this method does not include $\text{ReLU}(\cdot)$ before $\text{softmax}(\cdot)$ in Eq.~\eqref{eq:ccf_inference} since it did not show any benefit empirically unlike in other methods.

\subsection{LM fusion by updating both hidden and memory cell states} \label{point2}
In \textit{cold fusion}, the update of the hidden state output from the LSTM decoder in Eq.~\eqref{eq:cf_3} directly affects the final inference unlike \emph{cell control fusion 1}.
Therefore, this section combines the concepts of the \textit{cold fusion} and \emph{cell control fusion 1} and further proposes variants of novel fusion methods by extending the \textit{cell control fusion 1} with the above hidden state consideration.

First, we simply combine \textit{cell control fusion 1} in Section~\ref{point1} with \textit{cold fusion}. We call this scheme as \textit{cell control fusion 2}.
The detailed equations are
\begin{subequations} \label{ccf2}
\begin{equation} \label{eq:ccf2_logit_transf}
    \bm{h}_{t}^\text{LM} = \bm{W}_1 \bm{l}_{t}^\text{LM} + \bm{b}_1
\end{equation}
\begin{equation}
    \bm{g}_{t} ^{\text{cell}} = \sigma(\bm{W}_2[\bm{c}_t; \bm{h}_{t}^\text{LM}] + \bm{b}_2)
\end{equation}
\begin{equation}
    \bm{c}_{t}^\text{CCF2} = \bm{c}_{t} + \bm{g}_{t} ^{\text{cell}} \odot \bm{h}_{t}^\text{LM}
\end{equation}
\begin{equation} \label{eq:ccf2_dec}
    \bm{s}_{t+1}, \bm{c}_{t+1} = \text{LSTM}(\text{input}, \bm{s}_t, \bm{c}_t^\text{CCF2})
\end{equation}
\begin{equation} \label{eq:cf_gate}
    \bm{g}_{t} ^{\text{state}} = \sigma(\bm{W}_3[\bm{s}_t; \bm{h}_{t}^\text{LM}] + \bm{b}_3)
\end{equation}
\begin{equation} \label{eq:cf_state}
     \bm{s}_{t}^\text{CCF2} = [{{\bm{s}_{t}}};\bm{g}_t^{\text{state}}\odot \bm{h}_{t}^\text{LM}]
\end{equation}
\begin{equation} \label{eq:cf_inference}
     \hat{p}(y_{t}|y_{<t}, x) = \text{softmax}(\text{ReLU}(\bm{W}_4 \bm{s}_{t}^\text{CCF2} + \bm{b}_4))\;.
\end{equation}
\end{subequations}
Note the calculations of Eq.~\eqref{eq:ccf2_logit_transf},~\eqref{eq:cf_gate}-\eqref{eq:cf_inference} are exactly same as of Eq.~\eqref{eq:cf}, and calculations of Eq.~\eqref{eq:ccf2_logit_transf}-\eqref{eq:ccf2_dec} are same as of Eq.~\eqref{eq:ccf_logit_transf}-\eqref{eq:ccf_dec} other than $\text{tanh}(\cdot)$ used in Eq.~\eqref{eq:ccf_logit_transf}, which shows some effectiveness on our preliminary investigation.
However, this straightforward extension does not outperform both \textit{cell control fusion 1} and \textit{cold fusion}.

As a next variant, we apply a similar cell control update mechanism (Eq.~\eqref{eq:ccf2_dec}) to the seq2seq decoder hidden state $\bm{s}_t$ as well.
That is, the original hidden state, $\bm{s}_t$, is replaced by $\bm{s}_t^{\text{CCF3}}$ in the LSTM update Eq.~\eqref{ff_dec}, which is transformed from the fused state in \textit{cold fusion} to match dimension. $\bm{s}_t^{\text{CCF3}}$ is expected to have more information since it contains additional linguistic context obtained from an external LM. The Eq. \eqref{feedbackfusion} explains this fusion method more in detail. We call this type of fusion \textit{cell control fusion 3} in this paper.

\begin{subequations} \label{feedbackfusion}
\begin{equation}
    \bm{h}_{t}^\text{LM} = \tanh(\bm{W}_1 \bm{l}_{t}^\text{LM} + \bm{b}_1)
\end{equation}
\begin{equation} 
    \bm{g}_{t}^{\text{state}} = \sigma(\bm{W}_2 [\bm{s}_{t}; \bm{h}_{t}^\text{LM}] + \bm{b}_2)
\end{equation}
\begin{equation} \label{added_gatecell}
    \bm{g}_t^{\text{cell}} = \sigma(\bm{W}_3 [\bm{c}_{t}; \bm{h}_{t}^\text{LM}] + b_3)
\end{equation}
\begin{equation} \label{edited_fusedstate}
    \bm{s}_{t}^\text{CCF3} = \bm{W}_4 [\bm{s}_{t}; \bm{g}_{t}^{\text{state}}\odot \bm{h}_{t}^\text{LM}] + \bm{b}_4
\end{equation}
\begin{equation} \label{FF_cellupdate}
    \bm{c}_{t}^\text{CCF3} = \text{CellUpdate}(\bm{c}_{t}, \bm{g}_t^{\text{cell}}\odot \bm{h}_{t}^\text{LM})
\end{equation}
\begin{equation} \label{ff_dec}
    \bm{s}_{t+1}, \bm{c}_{t+1} = \text{LSTM}(\text{input}, \bm{s}_t^{\text{CCF3}}, \bm{c}_t^{\text{CCF3}})
\end{equation}
\begin{equation}
    \hat{p}(y_{t}| x,y_{<t}) = \text{softmax}(\text{ReLU}(\bm{W}_5 \bm{s}_{t}^\text{CCF3} + \bm{b}_5))
\end{equation}
\end{subequations}
For $\text{CellUpdate}$ function in Eq. \eqref{FF_cellupdate}, we compared two different calculations:
\begin{equation} \label{eq:CCF3_add}
    \bm{c}_{t} + \bm{g}_t^{\text{cell}}\odot \bm{h}_{t}^\text{LM}
\end{equation}
\begin{equation} \label{eq:CCF3_DNN}
    \bm{W}_0 [\bm{c}_{t};\bm{g}_t^{\text{cell}}\odot \bm{h}_{t}^\text{LM}] + \bm{b}_0
\end{equation}
In the case of Eq.~\eqref{eq:CCF3_DNN}, the affine transformation of $[\bm{c}_{t};\bm{g}_t^{\text{cell}}\odot \bm{h}_{t}^\text{LM}]$ would cause gradient vanishing problem in theory. However, we found that in practice, the method works best among all proposed methods.

\section{Experiments} \label{ses:exp}
We first compared all our proposed methods described in Section \ref{ses:proposed} with \textit{shallow fusion}, \textit{deep fusion}, and \textit{cold fusion} on the 100hrs subset of the Librispeech corpus~\cite{panayotov2015librispeech} as a preliminary experiment.
Then, we further investigate some selected methods with two other experiments: mono-lingual ASR setup on the Librispeech 960hrs corpus and a transfer learning setup from a multilingual ASR (MLASR) base model to a low-resourced language, Swahili in IARPA Babel~\cite{karafiat2016multilingual}.

We used \textit{shallow fusion} all the time in decoding phase for every trained model. For example, we can additionally use \textit{shallow fusion} decoding for the seq2seq model trained with a \textit{cold/cell-control fusion} scheme. We refer to~\cite{toshniwal2018comparisonSFnDFnCF} to justify the comparison between the baseline seq2seq model with \textit{shallow/deep fusion} decoding and the seq2seq model trained using \textit{cold/cell-control fusion} with \textit{shallow fusion} decoding. The baseline seq2seq model above means a seq2seq model trained without any fusion method.

All models are trained with joint CTC-attention objective as proposed in~\cite{watanabe2017attctc},
\begin{equation} \label{eq:joint_train}
 L_{\text{MTL}} = \alpha L_{\text{CTC}} + (1-\alpha) L_{\text{Attention}}
\end{equation}
where $L_{\text{CTC}}$ and $L_{\text{Attention}}$ are the losses for CTC and attention repectively, and $\alpha$ ranges between 0 and 1 inclusively. In decoding, we did attention/CTC joint decoding with RNNLM~\cite{hori2017multi}. 
In all the experiments, we represented each frame of 25ms windowed audio having 10ms shift by a vector of 83 dimensions, which consists of 80 Mel-filter bank coefficients and 3 pitch features. 
The features were normalized by the mean and the variance of the whole training set. All experiments were done based on ESPnet toolkit~\cite{watanabe2018espnet}.

For Librispeech, training and decoding configurations of the seq2seq model are shown in Table~\ref{tab:config_seq2seq}.
We trained both character-level and word-level RNNLMs on 10\% of the text publicly available for Librispeech\footnote{http://www.openslr.org/11/}, which is roughly 10 times the 960 hours of the transcriptions in terms of the data size. 

In the MLASR transfer learning scenario, the base MLASR model was trained exactly same as in ~\cite{cho2018multilingual}. The base model was trained using 10 selected Babel languages, which are roughly 640 hours of data: Cantonese, Bengali, Pashto, Turkish, Vietnamese,
Haitian, Tamil, Kurmanji, Tokpisin, and Georgian. The model parameters were then fine-tuned using all Swahili corpus in Babel, which is about 40 hours. During the transfer process, we used the same MLASR base model with three different ways: 2-stage transfer (see ~\cite{cho2018multilingual} for more details), \textit{cold fusion}, and \textit{cell control fusion 3 (affine)}. We included \textit{cold fusion} in this comparison since \textit{cold fusion} showed its effectiveness in domain adaptation in~\cite{sriram2017coldfusion}. The character-level RNNLM was trained on all transcriptions available for Swahili in IARPA Babel.
\vspace*{-5pt}
\begin{table}[h!]
\caption{Experiment details\label{tab:config_seq2seq}}
\begin{centering}
{\scriptsize{}}%
\begin{tabular}{cc}
\hline 
{\scriptsize{}Model Configuration} & \tabularnewline
\hline 
\hline 
{\scriptsize{}Encoder} & {\scriptsize{}Bi-LSTM}\tabularnewline
{\scriptsize{}\# encoder layers} & {\scriptsize{}8}\tabularnewline
{\scriptsize{}\# encoder units} & {\scriptsize{}320}\tabularnewline
{\scriptsize{}\# projection units} & {\scriptsize{}320}\tabularnewline
{\scriptsize{}Decoder} & {\scriptsize{}Bi-LSTM}\tabularnewline
{\scriptsize{}\# decoder layers} & {\scriptsize{}1}\tabularnewline
{\scriptsize{}\# decoder units} & {\scriptsize{}300}\tabularnewline
{\scriptsize{}Attention} & {\scriptsize{}Location-aware}\tabularnewline
\hline 
\hline 
{\scriptsize{} Training Configuration} & \tabularnewline
\hline 
\hline 
{\scriptsize{}Optimizer} & {\scriptsize{}AdaDelta~\cite{zeiler2012adadelta}}\tabularnewline
{\scriptsize{}Initial learning rate} & {\scriptsize{}1.0}\tabularnewline
{\scriptsize{}AdaDelta $\epsilon$} & {\scriptsize{}$1e^{-8}$}\tabularnewline
{\scriptsize{}AdaDelta $\epsilon$ decay} & {\scriptsize{}$1e^{-2}$}\tabularnewline
{\scriptsize{}Batch size} & {\scriptsize{}36}\tabularnewline
{\scriptsize{}ctc-loss weight ($\alpha$)} & {\scriptsize{}$0.5$}\tabularnewline
\hline 
\hline 
{\scriptsize{}Decoding Configuration} & \tabularnewline
\hline 
\hline 
{\scriptsize{}Beam size} & {\scriptsize{}20}\tabularnewline
{\scriptsize{}ctc-weight ($\lambda$~\cite{hori2017multi})} & {\scriptsize{}$0.3$}\tabularnewline
\hline 
\end{tabular}
\par\end{centering}{\scriptsize \par}
\end{table}
\subsection{Preliminary experiments: Librispeech 100hrs}
Table \ref{tab:SFnDFnCFnCCF} compares the proposed \textit{cell control fusion} methods to the conventional fusion methods. 
Both \textit{cold fusion} and \textit{cell control fusion 1} show similar performance, but the performance of \textit{cell control fusion 2} was degraded. This implies that simply combining \textit{cold fusion} and \textit{cell control fusion 1} does not have any benefit. Then,
The \textit{cell control fusion 3} methods, which extended \textit{cell control fusion 2} by applying cell control update mechanism (Eq.~\eqref{eq:ccf2_dec}) to the seq2seq decoder hidden state (Eq.~\eqref{ff_dec}), outperformed all the previous fusion methods in most cases. This suggests that applying the cell control update mechanism for both cell and hidden states consistently, further improves the performance. Among the two~\textit{cell control fusion 3} methods, \textit{cell control fusion 3 (affine)} outperformed all the other methods in every case. 
During the decoding, the shallow fusion parameter $\gamma$ in Eq.~\eqref{eq:shallow_comb} was set to 0.3.

\vspace*{-5pt}
\begin{table}[h!]
\caption{Comparison of previous and cell control fusion methods on Librispeech 100 hours: Character-level decoding (\%WER) \label{tab:SFnDFnCFnCCF}}
\centering{}{\scriptsize{}}%
\setlength\extrarowheight{-3pt}
\resizebox{0.48\textwidth}{!}{
\begin{tabular}{l||c|c|c|c}
\hline 
{\scriptsize{}Fusion} & {\scriptsize{}dev} & {\scriptsize{}test} & {\scriptsize{}dev} & {\scriptsize{}test}\tabularnewline
{\scriptsize{}method} & {\scriptsize{}clean} & {\scriptsize{}clean} & {\scriptsize{}other} & {\scriptsize{}other} \tabularnewline
\hline 

{\scriptsize{}Shallow fusion} & {\scriptsize{}16.9} & {\scriptsize{}16.7} & {\scriptsize{}45.6} & {\scriptsize{}47.9}\tabularnewline

{\scriptsize{}Deep fusion} & {\scriptsize{}17.1} & {\scriptsize{}17.0} & {\scriptsize{}45.9} & {\scriptsize{}48.3}\tabularnewline

{\scriptsize{}Cold fusion} & {\scriptsize{}16.7} & {\scriptsize{}16.4} & {\scriptsize{}45.5} & {\scriptsize{}47.8}\tabularnewline

{\scriptsize{}Cell control fusion 1} & {\scriptsize{}16.4} & {\scriptsize{}16.5} & {\scriptsize{}45.4} & {\scriptsize{}47.7}\tabularnewline

{\scriptsize{}Cell control fusion 2} & {\scriptsize{}17.4} & {\scriptsize{}16.8} & {\scriptsize{}45.9} & {\scriptsize{}48.4}\tabularnewline

{\scriptsize{Cell control fusion 3 (sum)}} & {\scriptsize{}16.7} & {\scriptsize{}16.3} & {\scriptsize{}45.3} & {\scriptsize{}47.2}\tabularnewline

{\scriptsize{}Cell control fusion 3 (affine)} & {\scriptsize{}\textbf{16.0}} & {\scriptsize{}\textbf{16.0}}& {\scriptsize{}\textbf{44.7}} & {\scriptsize{}\textbf{46.6}}\tabularnewline


\hline 
\end{tabular}{\scriptsize \par}
}
\end{table}

\subsection{Librispeech 960 hrs}
In this setting, we decoded in two ways: character-level decoding and multi-level (character and word) decoding~\cite{hori2017multi}. 
The word-level RNNLM used for the multi-level decoding has 20,000 as its vocabulary size. 
The results are shown in Table~\ref{tb:librispeech960_wer_charlevel} and~\ref{tb:librispeech960_wer_multilevel} respectively. 
For both cases, \textit{cell control fusion 3 (affine)}, performed the best followed by \textit{shallow fusion}, and \textit{cold fusion}. 
Also, we observed that the gap in the WER between \textit{cell control fusion 3 (affine)} and \textit{shallow fusion} is larger when we use multi-level decoding. This suggests that with the advanced decoding algorithm, the ~\textit{cell control fusion 3} benefits more in performance.
Note that $\gamma$ was set to 0.3 for character-level decoding and 0.5 for multi-level decoding.

\setlength{\tabcolsep}{6pt}
\vspace*{-5pt}
\begin{table}[h]
	\centering
		\caption{LibriSpeech 960 hours: Character-level decoding (\%WER)}
		\label{tb:librispeech960_wer_charlevel}
		\resizebox{0.48\textwidth}{!}{
		\begin{tabular}{l||c|c|c|c}
		     \hline
		     Fusion & dev & dev & test & test \\
		     method & clean & other & clean & other \\
		     \hline
		      Shallow fusion                    & 6.1 & 17.6 & 6.1 & 18.1 \\
		      Cold fusion             & 6.1 & 18.1 & 6.3 & 18.7 \\
		      Cell control fusion 3 (affine)     & \textbf{6.0} & \textbf{17.1} & \textbf{6.1} & \textbf{17.9} \\
		      \hline
		\end{tabular}}
	\vskip -3mm
\end{table}
\vspace*{-5pt}
\begin{table}[h]
	\centering
		\caption{LibriSpeech 960 hours: Multi-level decoding (\%WER)}
		\label{tb:librispeech960_wer_multilevel}
		\resizebox{0.48\textwidth}{!}{
		\begin{tabular}{l||c|c|c|c}
		     \hline
		     Fusion & dev & dev & test & test \\
		     method & clean & other & clean & other \\
		     \hline
		      Shallow fusion                    & 5.4 & 15.8 & 5.4 & 16.6 \\
		      Cold fusion             & 5.4 & 16.2 & 5.6 & 17.1 \\
		      Cell control fusion 3 (affine)     & \textbf{5.2} & \textbf{15.5} & \textbf{5.2} & \textbf{16.2} \\
		      \hline
		\end{tabular}}
	\vskip -3mm
\end{table}

\subsection{Transfer learning to low-resourced language}
Finally, Table~\ref{tb:MLASR_transfer} shows the result of the transfer learning to a low-resourced language (Swahili). The \textit{cold fusion} transfer improved the performance from simple 2-stage, showing its effectiveness in this scenario but \textit{cell control fusion 3 (affine)} improved the performance further. \textit{cell control fusion 3 (affine)} outperforms \textit{cold fusion} not only in this MLASR transfer learning setup but also in the previous mono-lingual ASR setup.
In decoding, $\gamma$ was set to 0.4.
\vspace*{-5pt}
\begin{table}[h]
	\centering
		\caption{Transfer learning from an MLASR base model to Swahili: Character-level decoding (\%CER, \%WER)}
		\label{tb:MLASR_transfer}{
		\begin{tabular}{lcc}
		     \hline
		     Fusion & eval set & eval set \\
		     method & \%CER & \%WER \\
		     \hline
		      Shallow fusion (2-stage transfer) \cite{cho2018multilingual}                   & 27.2 & 56.2 \\
		      Cold fusion             & 25.8 & 52.9 \\
		      cell control fusion 3 (affine)     & \textbf{24.5} & \textbf{50.7} \\
		      \hline
		\end{tabular}}
	\vskip -3mm
\end{table}

\section{Conclusion} \label{ses:conclusion}

In this paper, several methods were proposed to integrate a pre-trained RNNLM during a seq2seq model training. First, we used information from the LM output to update the memory cell in the seq2seq model decoder, which performed similar to \textit{cold fusion}. Then, we 
extended this model to additionally update the seq2seq model hidden state given the LM output. For the scheme, Several formulas were compared. 
Among the proposed methods, \textit{cell control fusion 3 (affine)} showed the best performance consistently on all experiments. 
In Librispeech, our best model improved WER by 3.7\%, 2.4\% for test clean, test other relatively to the shallow fusion baseline, with multi-level decoding. 
For the transfer learning setup from a MLASR base model to the IARPA Babel Swahili model, the best scheme improved the transferred model performance on eval set by 9.9\%, 9.8\% in CER, WER relatively to the 2-stage transfer baseline. In the future, we will explore how the best method performs by the amount of additional text data used for RNNLM training.
 \newpage

\bibliographystyle{IEEEbib}

\newpage
\section*{Important Notice}\label{sec:notice}
Dear Readers,\\[5pt]
Please be advised that the minor differences observed in the comparative performance reported in this paper may be attributed to variations in GPU specifications. The compute server used in the experiments included both Tesla K80 and GTX 1080 Ti GPUs, and jobs were randomly assigned to nodes without any deliberate selection based on GPU type. When I became aware of this issue, I attempted to withdraw the paper; however, the publication process had already been finalized. I had planned to address the matter by re-running the experiments using a fixed GPU configuration; however, I was unable to directly implement this plan. Unfortunately, due to significant changes in the computing environment and the elapsed time since the original experiments, replicating the exact conditions is no longer feasible.\\[5pt]
I sincerely apologize for any confusion this may have caused. Please consider this information when interpreting our results, and feel free to cite the paper at your discretion.\\[5pt]
For clarity, the first author, Jaejin Cho, takes full responsibility for the matters described above. This statement is made in the spirit of academic integrity and transparency, and does not imply legal liability.\\[5pt]
Thank you for your understanding.\\[5pt]
Sincerely,\\
Jaejin Cho

\end{document}